\documentclass[12pt,preprint]{aastex}

\shorttitle{31/250~GHz observations of the Helix}
\shortauthors{Casassus et al.}

\begin{document}


\title{Anomalous radio emission from dust in the Helix}

\author{S. Casassus\altaffilmark{1}, A.C.S.~Readhead\altaffilmark{2},
  T.J.~Pearson\altaffilmark{2}, L.-\AA.~Nyman\altaffilmark{3}, \\
  M.C.~Shepherd\altaffilmark{2}, L.~Bronfman\altaffilmark{1}}

\altaffiltext{1}{Departamento de Astronom\'{\i}a, Universidad de Chile,
Santiago, Casilla 36-D, Chile}
\altaffiltext{2}{Owens Valley Radio Observatory, California Institute
  of Technology, Pasadena, CA 91125}
\altaffiltext{3}{Onsala Space Observatory, 439\,92 Onsala, Sweden}

\begin{abstract}
A byproduct of experiments designed to map the cosmic microwave
background is the recent detection of a new component of foreground
Galactic emission. The anomalous foreground at $\sim$10--30~GHz,
unexplained by traditional emission mechanisms, correlates with
100$\mu$m dust emission.  We use planetary nebulae (PNe) as
astrophysical laboratories to test known radio emission processes, and
report that in the Helix the emission at 31~GHz and 100~$\mu$m are
well correlated, and exhibit similar features on sky images, which are
absent in H$\beta$.  Upper limits on the 250~GHz continuum emission in
the Helix rule out cold grains as candidates for the 31~GHz emission,
and provide spectroscopic evidence for an excess at 31~GHz over
bremsstrahlung. We estimate that the 100~$\mu$m-correlated radio
emission, presumably due to dust, accounts for at least 20\% of the
31~GHz emission in the Helix. This result strengthens previous
tentative interpretations of diffuse interstellar medium spectra
involving a new dust emission mechanism at radio frequencies.  Very
small grains, thought not to survive in evolved PNe, have not been
detected in the Helix, which hampers interpreting the new component in
terms of electric dipole emission from spinning grains. The observed
iron depletion in the Helix favors considering the identity of this
new component to be magnetic dipole emission from hot ferromagnetic
grains. The reduced level of free-free continuum we report also
implies an electronic temperature of $T_e = 4600\pm1200~$K for the
free-free emitting material, which is significantly lower than the
temperature of $9500\pm500$~K inferred from collisionally-excited
lines.

\end{abstract}

\section{Introduction}\label{sec:intro}

Continuum emission mechanisms from ionized nebulae or from the diffuse
interstellar medium (ISM) at radio frequencies (lower than
$\sim$90~GHz), have up to now been attributed to two components:
Free-free emission (thermal bremsstrahlung), or synchrotron emission.
But the problem of separating Galactic foregrounds from cosmic
microwave background (CMB) radiation has motivated a careful review of
the emission mechanisms from the ISM.  Dust-correlated radio emission,
with hints of anomalous spectral properties, has first been reported
in the analysis of {\em COBE} data \citep{kog96}.  The `anomalous
foreground' detected at 14~GHz by \citet{lei97} correlates with {\em
IRAS} 100$\mu$m maps, and has a flat spectral index characteristic of
free-free emission, but the corresponding H$\alpha$ emission is
absent.  \citet{dl98a} rule out, on energetic grounds, hot plasma
interpretations of the anomalous foreground, and propose a model in
which it is due to electric dipole radiation from spinning very small
dust grains (VSGs). Their proposition has found verification from
statistical evidence brought forward by \citet{deol99,deol02}.

\citet{fin02} have detected a spectral energy distribution (SED) that
is inconsistent with free-free emission for the 100~$\mu$m-correlated
radio emission in the diffuse H\,{\sc ii} region LPH~201.663+1.643
(LPH~201.6 hereafter). But the interpretation of the SED of LPH~201.6
in terms of the \citet{dl98a} models meets two difficulties. One is
the premise that radio emission in LPH~201.6 is proportional to far-IR
dust emission. Another difficulty is that the positive 5-10~GHz
spectral index reported by Finkbeiner et al. could also be accounted
for by unresolved optically thick emission \citep{cul02}.

It would seem that apart from well-studied free-free and synchrotron
emission, there is an additional component in the SED of the diffuse
ISM.  Are there ways to constrain the many free parameters in the
\citet{dl98b} model, such as the precise identity of the spinning
VSGs?  Is spinning dust ubiquitous in the Galaxy? Is it important in
objects such as planetary nebulae (PNe), or H\,{\sc ii} regions? What
is the importance of magnetic dipole emission from classical grains,
also proposed by \citet{dl99}? Improved knowledge of these new
emission mechanisms is crucial to the interpretation of
radio-continuum observations as a diagnostic of physical conditions.


Planetary nebulae are perhaps the simplest of ionized nebulae: an
ionized expanding envelope around an exposed stellar core. Their being
bright and isolated objects and their relative simplicity compared to
star-forming regions has given PNe a central role in the development
of nebular astrophysics \citep{ost89}. Evolved PNe, and in particular
NGC~7293 (the Helix), are important for the study of the late stages
of PN evolution and feed-back into the ISM. The large
angular size of the Helix, about 10~arcmin in diameter, stems from its
proximity, at a distance of $\sim$200~pc \citep{har97}, but also from
its huge physical size compared to other PNe, with emission traced at
diameters of up to 1~pc \citep{spe02}. The large sizes and ages of
evolved PNe make them useful probes of nebular structure, and of dust
grain survival in the PN phase. But it is their very size which makes
their observation difficult, especially at radio frequencies.

Resolved radio-frequency images of the Helix are
scarce. \citet{zij89}, as part of an imaging survey of PNe with the
VLA, reported a 14.9~GHz map of the Helix. The NRAO VLA Sky Survey
\citep[NVSS]{con298} detected the brightest clumps of the nebular
ring, but a more sensitive 1.4~GHz image was obtained by
\citet{rod02}. These three images are heavily affected by flux losses:
because of incomplete sampling in the $uv$ plane the reconstructed
images have missing spatial frequencies, and part of the extended
nebular emission is lost. The 1.4~GHz high resolution images reveal
the presence of bright background sources within the nebular diameter,
which distort low resolution images and spoil integrated nebular flux
density measurements at low frequencies.

Here we report evidence that 31~GHz emission in the Helix is due not
only to free-free emission, as currently thought, but also to a new
dust grain emission mechanism, other than the traditional thermal
emission from vibrations of the charge distribution \citep{dl99}. Part
of the 31~GHz emission shares similar characteristics with the
anomalous CMB foreground due to the diffuse ISM.  The discovery of a
new radio dust emission mechanism is the first in a specific object as
well studied as the Helix, and strengthens the tentative
interpretation of the SED of LPH~201.6 proposed by \citet{fin02}.

We have used the Cosmic Background Imager (CBI) to map the Helix at
31~GHz, which allows us to present integrated flux densities and
spatially resolved images at low resolution, but with little flux
loss. We find the background sources visible in the 1.4~GHz images of
the Helix are negligible at 31~GHz.  In addition we have used the Sest
IMaging Bolometer Array \citep[SIMBA]{nym01}, at the Swedish-ESO
Submillimetre Telescope (SEST) to search for possible cold dust
emission at 250~GHz, and to obtain a comparison flux density for the
31~GHz data, to explore the SED of evolved PNe. The Helix is not
detected in the SIMBA image, but the observations allow us to put
upper limits on the nebular flux at 250~GHz. The SIMBA observations
are complemented by weaker upper limits from the {\em WMAP} first-year
data release \citep{ben03}.

The CBI observations of the Helix, which constitute the first maps of
the anomalous dust-correlated emission, are presented in
Section~\ref{sec:cbi}, along with a discussion of flux losses and
background source contamination. The SIMBA upper limits on the radio
continuum at 250~GHz are given in Sec.~\ref{sec:simba}. Our results
are discussed in the context of the SED of the Helix in
Sec.~\ref{sec:disc}, in which we estimate the ionized mass of the
Helix to be 0.096~M$_\odot$, with a 30\% accuracy. Sec.~\ref{sec:conc}
concludes. We use the Perl Data Language ({\tt http://pdl.perl.org})
for all data analysis, unless otherwise stated.

\section{CBI observations} \label{sec:cbi}

The CBI \citep{pad02} is a planar interferometer array with 13
antennas, each 0.9~m in diameter, mounted on a 6~m tracking platform,
which rotates in parallactic angle to provide uniform
$uv$-coverage. It is located in Chajnantor, Atacama, Chile. The CBI
receivers operate in 10 frequency channels, with 1~GHz bandwidth each,
giving a total bandwidth of 26--36~GHz.

We have observed the Helix (J2000 RA=22:29:38, DEC= --20:50:13) in two
different array configurations, resulting in different full-widths at
half maximum (FWHM) for the Gaussian fits to the synthesized beam: we
have integrated for 4800s on 2001-Sep-7 and Nov-8, with an elliptical
beam of 4.90$\times$4.52~arcmin$^{2}$, and for 8250s on
2002-Nov-17,18,21,22, with a beam of 10.1$\times$8.19~arcmin$^{2}$.
Ground emission (spill-over), or the Moon, can enter in the side-lobes
of the receivers, and a comparison field, observed at the same
elevation as the target, is required for ground or Moon
cancellation. This reference field is offset by 8~arcmin in RA,
trailing the target, and is observed for the same integration time as
the object field. 

Flux calibration is performed using Tau\,A and Saturn, and is
described in detail by \citet{mas03}. The internal calibration
consistency on the CBI is better than 1\%, and the flux density scale,
originally set by absolute calibrations at the Owens Valley Radio
Observatory with an uncertainty of 3.3\% \citep{mas03}, has now been
improved by comparison with the WMAP temperature measurement of
Jupiter \citep{pag03} to an uncertainty of 1.3\% \citep{rea03}, which
is used in this work.

\subsection{Observed 31~GHz flux densities} \label{sec:cbiflux}

To extract the nebular flux density in a photometric aperture, we
reconstruct a clean image of the Helix by fitting an image-plane model
to the observed visibilities. The clean image is restored by adding
the residual dirty map to the model, convolved with the best-fit
elliptical beam. Using the `modelfit' task of the difmap package
\citep{she97}, we find a two component model gives a very good fit to
the data: a large elliptical Gaussian roughly following the optical
nebular ring, with a negative point source inset in the larger
component to account for the central plateau obvious in visible images
of NGC~7293 \citep[for instance]{ode98}.  For the purpose of
extracting an integrated nebular flux density, we prefer the use of
natural weights, which give the lowest noise in the restored image.

The flux density in a circular photometric aperture encompassing the
whole nebula is 992$\pm$35~mJy in the 2001 data.  The aperture radius
is set at 15~arcmin, the value beyond which the flux increment is less
than its rms uncertainty.  For comparison, the flux density in a
12~arcmin aperture is 934$\pm$28~mJy.  We use the 2002 data for a
counter check, and obtain 1000$\pm$31~mJy in an optimal aperture of
17~arcmin, or 966$\pm$26~mJy in 15~arcmin. Thus our best value for the
integrated 31~GHz flux density is 996$\pm$21, obtained by averaging
the 2001 and 2002 data (note this is without flux-loss correction, see
below).

We estimate the measurement uncertainty by multiplying the noise level
in Jy/pixel by $\sqrt{N\,N_\mathrm{beam}}$, where $N$ is the number of
pixels within the aperture, and $N_\mathrm{beam}$ is taken to
represent the number of correlated pixels (those that fall in one
beam). The 10 channel information over 26--36~GHz only places weak
constraints on the spectral index: a fit in the form $\kappa
(\nu/31.5)^\alpha$, where $\nu$ is the channel frequency in GHz, gives a
best fit $\kappa = 1.01\pm0.11$~Jy and $\alpha =0.17^{+1.19}_{-1.08}$,
at 68.3\% confidence
(i.e. $\sim\pm1\sigma$).


%
%

\subsection{Background source contamination at 31~GHz}

Background sources show up distinctively at 1.4~GHz, in Fig.~1 of
\citet{rod02}, and are a problem when reporting integrated fluxes for
the Helix.  But we do not detect them in our 31~GHz images, presumably
because they have sufficiently negative spectral indexes.

\begin{figure}
\epsscale{0.5} 
\plotone{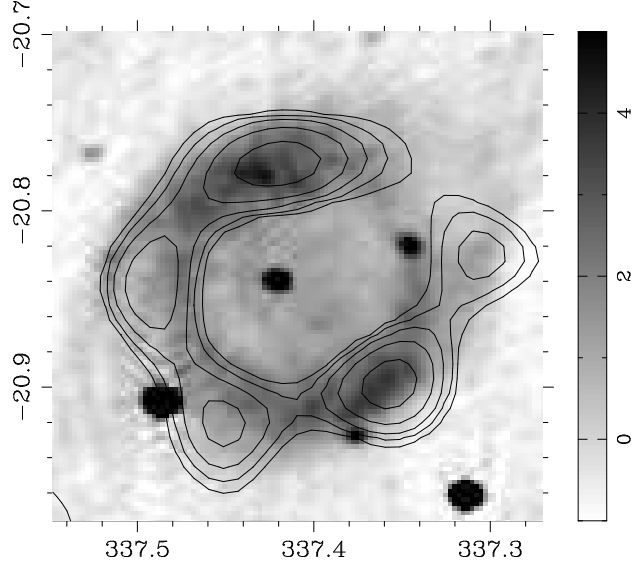}
\caption{The MEM model of the 31~GHz CBI visibilities of the Helix in
  contours, overlaid on a grey scale image at 1.4~GHz of the same
  region \citep{rod02}. All contour levels are increasing. The
  $x$-axis and $y-$axis are J2000 right ascension (RA), and
  declination (DEC), both in degrees. \notetoeditor{This figure should
  be adjusted in size so as to fit in one column.}
  \label{fig:nvss_mem}}
\end{figure}

To compare with the 1.4~GHz image of the Helix provided by
L.F.~Rodriguez, we extract the best resolution possible from our
31~GHz visibility data by using the MEM algorithm in AIPS++,
restricting to visibilities with $uv$-radii greater than 300. An
overlay of this image-plane MEM model with the 1.4~GHz image of the
Helix, shown in Fig.~\ref{fig:nvss_mem}, demonstrates that none of the
1.4~GHz point sources is present at 31~GHz, at least at a level
comparable to the nebular specific intensities.


The brightest point source at 1.4~GHz is  J2229-2054, with
$\sim0.1~$Jy in the NVSS image, and 0.15~Jy in \citet{rod02}, and is
12 times brighter than the next brightest point source within the
nebular diameter. As another test on the intensity of the 1.4~GHz
point sources in the CBI image, we add to the parametrized image-plane
model of the observed visibilities a single point source component,
with position fixed at that of  J2229-2054, but with a variable flux
density.  The best-fit model to the CBI visibilities, obtained with
`modelfit', gives a flux density of 12~mJy. Thus  J2229-2054, if
present at 31~GHz, is barely above the noise of the CBI image of
9~mJy/beam.

The black dot on the CBI image in Fig.\ref{fig:imhelixunf} lies at the
location of  J2229-2054, and is drawn to highlight its absence.

\subsection{CBI 31~GHz image and comparison with H$\alpha$, H$\beta$, 60~$\mu$m
   100~$\mu$m and 180~$\mu$m images.} \label{sec:images}

To guide the interpretation of the 31~GHz data, we compare with
simulations of the CBI observations on H$\alpha$, H$\beta$, 60$\mu$m
and 100$\mu$m template images.  We use as input images the
continuum-subtracted (CC) Southern H-Alpha Sky Survey Atlas
\citep[SHASSA]{gau01}, the H$\beta$ image\footnote{obtained from
Romano Corradi's web page {\tt \small
www.ing.iac.es/$\sim$rcorradi/HALOES/}} by \citet{ode98}, the
HIRES-reprocessed {\em IRAS} 60~$\mu$m and 100~$\mu$m
images\footnote{with default options, see {\tt \small
irsa.ipac.caltech.edu/IRASdocs/hires\_proc.html}}, and the {\em
ISO}\footnote{ISO is an ESA project with instruments funded by ESA
Member States (especially the PI countries: France, Germany, the
Netherlands and the United Kingdom) and with the participation of ISAS
and NASA. The ISO TDT and AOT codes for the image used here are
16701206 and PHT22, and the observer is P.~Cox. Note the calibrated
180~$\mu$m image is affected by a constant offset, by about the same
amount as the peak nebular intensity, which probably reflects
calibration problems.} 180~$\mu$m image of the Helix obtained with the
ISOPHOT camera \citep{lem96}.

Simulation of the CBI observations is performed with the MockCBI
program (Pearson 2000, private communication), which calculates the
visibilities $V(u,v)$ on the input images $I_\nu(x,y)$ with the same
$uv$ sampling as a reference visibility dataset:
\begin{equation} V(u,v) =
\int_{-\infty}^{\infty} A_\nu(x,y) I_\nu(x,y)\exp\left[2\pi i
(ux+vy)\right] \frac{dx\,dy}{\sqrt{1-x^2-y^2}}, 
\end{equation}
where $A_\nu(x,y)$ is the CBI primary beam and $x$ and $y$ are the
direction cosines relative to the phase center in two orthogonal
directions on the sky.  Thus MockCBI creates the visibility dataset
that would have been obtained had the sky emission followed the
template. In the case of the H$\beta$ input image, care is taken to
remove bright stars. The template images have much finer resolution
than the CBI synthesized beam, so that their final resolution is
essentially the same as that of the CBI images.

\begin{figure}
\epsscale{1.}  \plotone{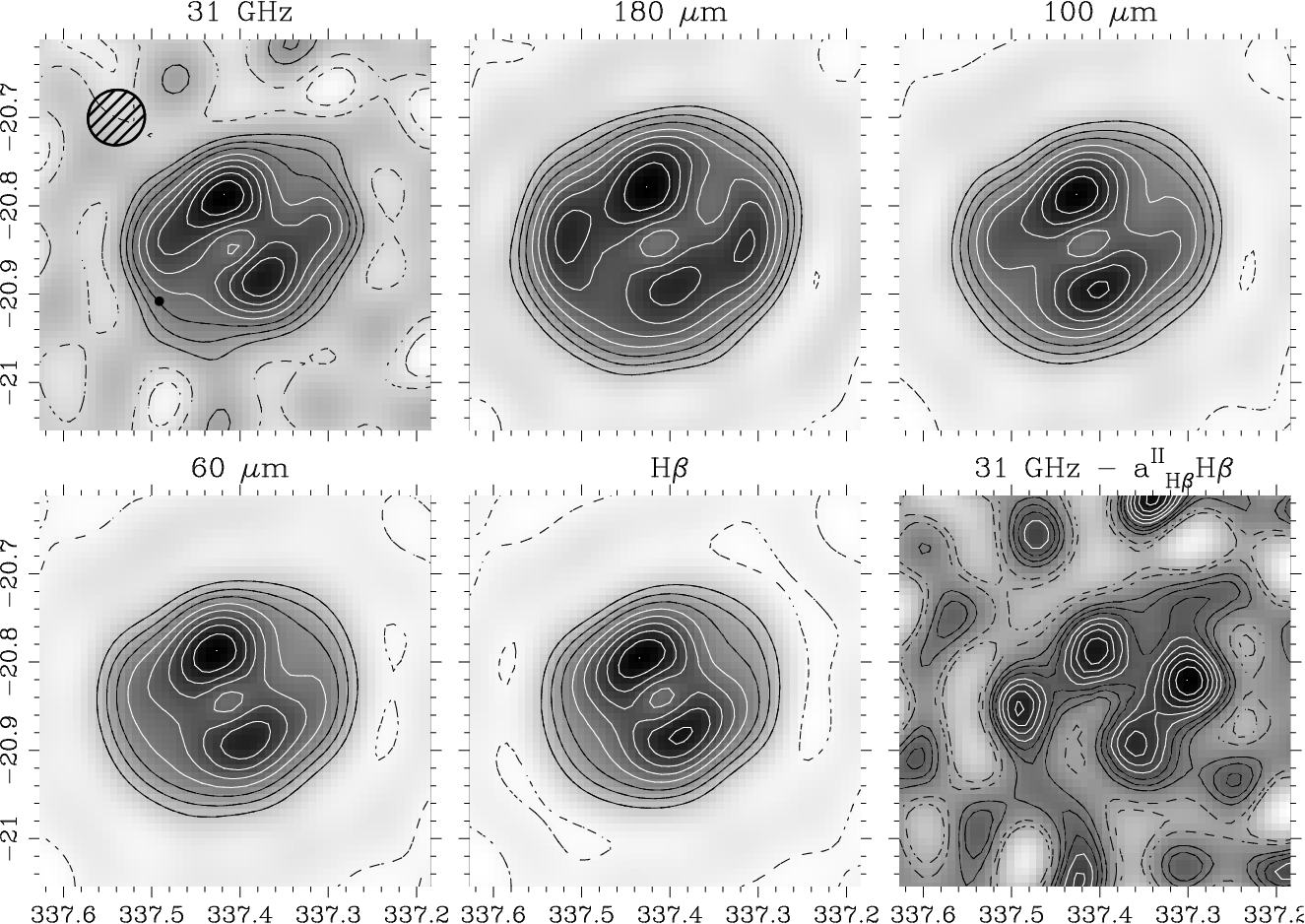}
\caption{The Helix at 31~GHz, 100~$\mu$m, 180~$\mu$m, 60~$\mu$m, and
  H$\beta$, as seen by the CBI: CBI observations have been simulated
  on the far-IR and H$\beta$ image to enable a faithful
  comparison. The image on the lower right results after subtracting
  the appropriately scaled H$\beta$ template from the CBI image (see
  Sec.~\ref{sec:crosscor}), and shows the anomalous emission.  Solid
  contour levels are linearly spaced in steps of 10\% the maximum
  intensity, and start at 10\%, and are overlaid for clarity on a
  linear gray scale image across the full range of intensity. Dotted
  contours are at the --10\% and zero levels.  The beam ellipse shown
  on the 31~GHz image, with a size of 3.89$\times$3.74~arcmin$^2$, is
  the same for all images. The noise level and peak specific intensity
  are at 8.9~mJy/beam and 0.135~Jy/beam for the 31~GHz data. The black
  dot on the CBI image lies at the location of J2229-2054.
\label{fig:imhelixunf}}
\end{figure}

The restored images are shown in Fig.~\ref{fig:imhelixunf}.  We use
the fitting of model components, as explained in
Sec.\ref{sec:cbiflux}, but this time restoring with uniform weights,
with an additional radial weight to improve resolution.  The 31~GHz
data faithfully reflects none of the comparison images. Rather, it
seems to correspond to a combination of all, with a stronger
resemblance to the 180~$\mu$m and 100~$\mu$m image than to the
60$\mu$m and H$\beta$ images. The 31~GHz emission is localized in two
bright North and South clumps (N-S clumps), each with East and West
extensions (E-W lobes).  The N-S clumps at 31~GHz reflect the H$\beta$
image, and are slightly offset from their far-IR counterparts. But the
E-W lobes are absent in H$\beta$.  The differences between H$\beta$
and 31~GHz are best appreciated in the lower-right panel of
Fig.~\ref{fig:imhelixunf}, where we have subtracted from the 31~GHz
image the H$\beta$ template scaled to the expected level of free-free
emission (see below). The remainder of this subtraction is the
anomalous emission: It cannot be accounted for by free-free emission.

\begin{figure}
\epsscale{0.5}
\plotone{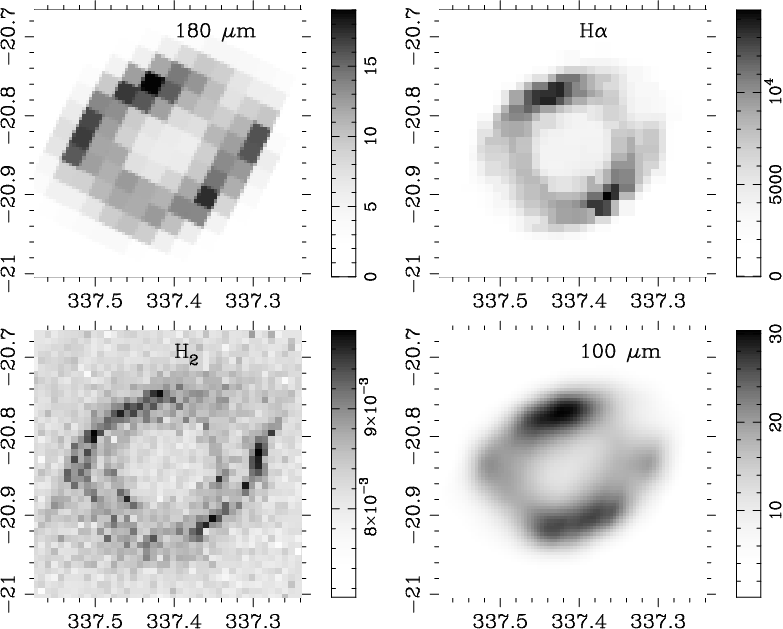}
\caption{Comparison images at 180~$\mu$m, H$\alpha$, H$_2$ (after
  removing the two brightest stars in the field), and 100~$\mu$m,
  highlighting the similarities between the far-IR images, and their
  differences with H$\alpha$.
  \label{fig:templates}}
\end{figure}

It can be noted that the anomalous emission map and the {\em ISO}
180~$\mu$m maps share similar morphologies, although the later is more
extended. The 100~$\mu$m and 60~$\mu$m are also more compact than the
180~$\mu$m map, as expected from grain heating due to central star
radiation, but the E-W lobes do not show as markedly as in the
180~$\mu$m map. The anomalous emission isn't as sharply confined to
the N-S clumps, and has a very bright Western lobe relative to the
rest of the nebula.  The only other map with such a bright Western
lobe is the ISOCAM 5--8.5~$\mu$m map from \citet[they refer to the
Western lobe as the Western rim]{cox98}, which is dominated by H$_2$
line emission (see Fig.~\ref{fig:templates}).

The H$\alpha$ simulated images with the SHASSA templates have
essentially the same morphology as the H$\beta$ simulation, which
serves as cross check. We prefer the H$\beta$ data since [N\,{\sc
ii}], which arises mostly in the nebular ring, contaminates the SHASSA
narrow band filter up to 50\% of the total flux.  The [N\,{\sc ii}]
contamination is reflected in the H$\alpha$ flux from the SHASSA
image, which is twice the value from \citet{ode98} (scaled for the
Balmer decrement): Transmission for the two [N\,{\sc ii}] lines is
about 50\% while their flux is about twice the H$\alpha$ flux.

To illustrate that the morphological trends shown in
Fig.~\ref{fig:imhelixunf} are not artifacts of the deconvolutions, we
present in Fig.~\ref{fig:templates} a comparison of the ISOPHOT
180~$\mu$m, the HIRES 100~$\mu$m images, the ISOCAM LW2 map\footnote{
ISO TDT 16701502}, and the SHASSA CC image. The SHASSA image is better
suited than the H$\beta$ image for comparison with the far-IR data
because it is closer in resolution, and because most stars are removed
in the continuum subtraction.  The morphological differences between
dust and H$\alpha$ are apparent, while the 180~$\mu$m-100~$\mu$m
similarity renders improbable their contamination from emission lines.



%
%

\subsection{Flux loss estimates} \label{sec:fluxloss}

The Helix is extended and resolved by the CBI, and there is bound to
be a measure of flux loss derived from missing low spatial frequencies
in the CBI's $uv$ coverage. We now estimate the amount of flux loss
with the template images mentioned in Sec.~\ref{sec:images}.  We
simulate the 2001 CBI observations using MockCBI, and after
reconstructing with natural weights, in the same manner as described
in Sec.~\ref{sec:cbiflux} for the CBI data, we extract the integrated
nebular flux density and compare with the input map. This strategy
also allows estimating systematic errors on the integrated flux
derived from the image reconstruction.

In the case of the 100$\mu$m image, the flux density in a circular
aperture of 15~arcmin in radius, as for the CBI photometry, dropped
from 305.8~Jy to 272.3~Jy, which means 89.0\% of the flux density was
recovered. We repeated the same exercise on the continuum-corrected
SHASSA image, recovering 92.0\% of the flux, and 90.0\% in the case of
the H$\beta$ image. The finest resolution is that of the H$\beta$
image, which probably best reflects the flux lost.

Therefore we estimate that 90\% of the flux is recovered by the CBI in
the 15$'$ circular photometric aperture we are using, and the actual
flux density of the Helix at 31~GHz is 1096~mJy, with a $\sim$10\%
uncertainty.

\subsection{Cross-correlations between CBI and template visibilities}
\label{sec:crosscor}

Here we cross-correlate the CBI data with the comparison templates, in
order to quantify the resemblance of the 31~GHz image to the {\em
IRAS} data. The 100~$\mu$m correlation is particularly interesting
because it is a comparison point with previous work on the anomalous
foreground.

We investigate a model in which the 31~GHz specific intensities,
$\vec{y}$, result from a linear combination of various components, as
in \citet{deol02} and \citet{fin02}: $\vec{y} = \sum_{i=1}^N a^{N}_i
\vec{x_i}$, where $N$ is the number of templates $\vec{x_i}$.  We fit
for conversion factors, with $N=1$, $\vec{y} =
a^{I}_{100\mu\mathrm{m}} \vec{x}_{100\mu\mathrm{m}}$ and $\vec{y} =
a^{I}_{\mathrm{H}\beta} \vec{x}_{\mathrm{H}\beta}$, and for a linear
combination of the H$\beta$ and 100$\mu$m templates, $N=2$, $\vec{y} =
a^{II}_{\mathrm{H}\beta} \vec{x}_{\mathrm{H}\beta} +
a^{II}_{100\mu\mathrm{m}} \vec{x}_{100\mu\mathrm{m}}$.  We also try
fitting for a general cross-correlation, with all templates, $N=4$:
$\vec{y} = \sum_i a^{IV}_i \vec{x_i} = a^{IV}_{\mathrm{H}\beta}
\vec{x}_{\mathrm{H}\beta} + a^{IV}_{100\mu\mathrm{m}}
\vec{x}_{100\mu\mathrm{m}} + a^{IV}_{60\mu\mathrm{m}}
\vec{x}_{60\mu\mathrm{m}} + a^{IV}_{180\mu\mathrm{m}}
\vec{x}_{180\mu\mathrm{m}}$. In our notation the roman number
superscripts on the linear coefficients refer to $N$, the number of
templates used in the fits.

\begin{figure}
\epsscale{1.}
\plottwo{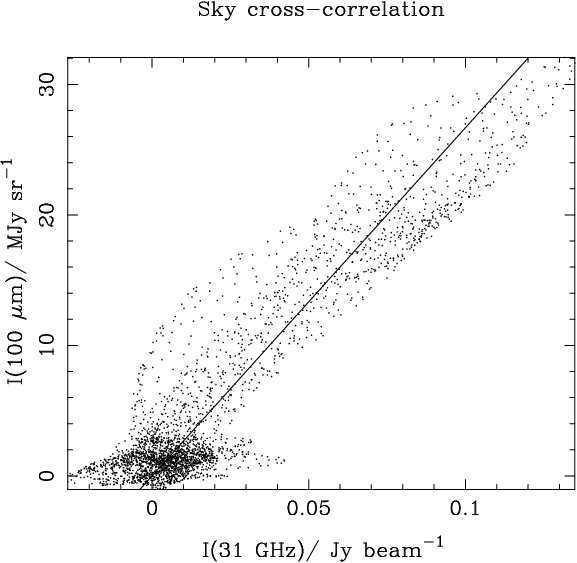}{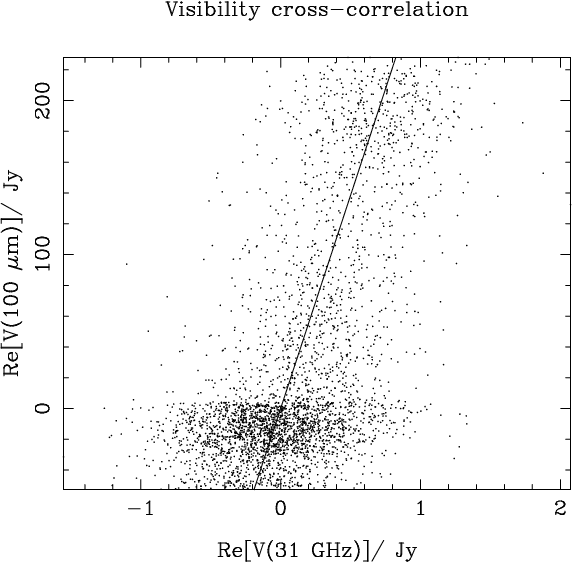}
\caption{Sky plane (left) and frequency plane (right)
  100~$\mu$m--31~GHz correlations. The straight line fits have
  dimensionless slopes of $3.45(2)~10^{-3}$ (sky) and
  $3.61(7)~10^{-3}$ (visibilities).  \label{fig:skycross}}
\end{figure}

Because of the linearity of the Fourier transform, the same
relationships hold for the complex visibilities and specific
intensities, provided the relationships are realistic. We performed
the cross-correlations both on the sky images and in frequency space,
shown in Fig.~\ref{fig:skycross}, confirming the results within the
uncertainties, which adds confidence in the linear combinations.
Proportionality between 31~GHz and the templates might not hold, for
instance, in the case of free-free emission, which is proportional to
the emission measure, and far-IR dust emission, which is proportional
to the column of dust.  One justification for a 31~GHz-100~$\mu$m
conversion factor might be that the diffuse UV nebular field makes an
important contribution to grain heating, another that dust emission is
important at 31~GHz.

The cross-correlation results are shown in
Table~\ref{table:crosscor}. It can be appreciated from the two
component fit ($a^{II}_{\mathrm{H}\beta}$ and
$a^{II}_{100\mu\mathrm{m}}$) that part of the 31~GHz visibilities
correlate with the 100~$\mu$m visibilities, and part with H$\beta$. No
significant correlation exists with 60~$\mu$m and 180~$\mu$m. We found
no trends with baseline length within the uncertainties.

The significance of the linear combination fits are difficult to
assess because the template images are smoothed versions of the
original maps, and their noise is undetermined. But we can use as
guideline the reduced $\chi^2$ in the absence of noise in the
templates, which are $\chi^2_I(100\mu\mathrm{m}) = 1.0971$,
$\chi^2_I(\mathrm{H}\beta) =1.0836$, $\chi^2_{II} = 1.0829$ and
$\chi^2_{IV} = 1.0821$. If the noise level in the templates is at
$10^{-3}$ the rms visibility values, then $\chi^2_I(\mathrm{H}\beta)
=1.0836 $, $\chi^2_I(100\mu\mathrm{m}) = 1.0609$, $\chi^2_{II} =
1.0280$ and $\chi^2_{IV} = 0.9897$. Thus the best fit model seems to
be model $IV$, then model $II$, followed by model $I$ with the
100~$\mu$m template if noise is included.

\begin{deluxetable}{rrrr}   
\tabletypesize{\small} \tablecaption{Cross-correlation results, given
in the format $a_i\pm\sigma(a_i)\left[
a_i/\sigma(a_i)\right]$. H$\beta$ entries have units of $10^{12}$ Jy /
W~m$^{-2}$, and all others have dimensionless units of $10^{-3}$. The
cross-correlations have been performed on the real part of the
visibilities, for the full $uv$-range of 88.4 -- 622~rad$^{-1}$. Note
we convert between flux densities, not brightness
temperatures. \label{table:crosscor}} \tablewidth{0pt}
\startdata 
\tableline\tableline \\
  $a^{I}_{\mathrm{H}\beta}$   & $a^{I}_{100\mu\mathrm{m}}$ &  $a^{II}_{\mathrm{H}\beta}$ &  $a^{II}_{100\mu\mathrm{m}}$ \\
  $ 2.65\pm 0.05[55]$ &  $ 3.61\pm 0.07[55]$   & $ 2.12\pm 0.29[7.2]$  & $ 0.75\pm 0.40[1.9]$ \\[0.5em]    
\tableline \\
 $a^{IV}_{\mathrm{H}\beta}$   & $a^{IV}_{100\mu\mathrm{m}}$ &  $a^{IV}_{60\mu\mathrm{m}}$  &  $a^{IV}_{180\mu\mathrm{m}}$  \\[0.2em]
  $ 3.05\pm 0.84[ 3.6]$  & $ 3.49\pm 1.59[ 2.2]$  & $-7.18\pm 3.38[-2.1]$  & $0.06\pm 1.26[0.0]$  \\
\enddata
\end{deluxetable}

Our 100$\mu$m--31~GHz conversion factor $a^I_{100\mu\mathrm{m}} =
3.61(7)~10^{-3}$ is close in value to that reported by \citet{fin02}
for the conversion from 100$\mu$m to 10~GHz.  After scaling units (our
$a^{I}_{100\mu\mathrm{m}}$ is dimensionless), the value for LPH~201.6
from \citet{fin02} is 3.84~10$^{-3}$ at 10~GHz. However, we caution
that \citet{fin02} compared with the 100~$\mu$m map from
\citet{sch98}, which cannot be used for the Helix because of
insufficient angular resolution.  The \citet{sch98} 100~$\mu$m flux
density is $\sim$30\% higher than the HIRES-reprocessed IRAS flux
density, both for the Helix and LPH~201.6. Had we used an integrated
flux density conversion factor and the \citet{sch98} 100~$\mu$m flux
density, our value for $a^I_{100\mu\mathrm{m}}$ would have been 30\%
lower.

On the other hand our value for $a^I_{100\mu\mathrm{m}}$ differs from
that reported by \citet{deol02} for Galactic latitudes $|b|>20^\circ$,
which is 2.2~10$^{-4}$ at 10~GHz and 3.6~10$^{-4}$ at 15~GHz.  In
\citet{deol02} the 15~GHz diffuse Galactic emission is modeled by a
three component linear fit, with H$\alpha$, 100~$\mu$m and synchrotron
templates. The synchrotron component decreases significantly from
10~GHz to 15~GHz, so if it is negligible at 15~GHz, their 100~$\mu$m
coefficient could perhaps be compared to our value for
$a^{II}_{100\mu\mathrm{m}}=0.75(40)~10^{-3}$, which agrees within the
uncertainties.

It must be noted that the conversion factors from \citet{deol02},
\citet{fin02}, and our work are not strictly comparable because they
correspond to different angular scales, and different objects. We
estimate the 12$'$-chopped observations from \citet{fin02} are
sensitive to spatial frequencies ($uv$-range) in the range
$\sim$200--400~rad$^{-1}$, similar to the CBI's (which roughly
corresponds to angular scales of $\sim$0.2~deg), while the data used
by \citet{deol02} are sensitive to multipoles lower than 30, roughly
equivalent to angular scales of 12~deg.

The ratio $a^{II}_{\mathrm{H}\beta}/a^{I}_{\mathrm{H}\beta} = 0.8$ can
be used as an upper limit to the fractional level of free-free
emission at 31~GHz: the single-component fit gives the conversion
factor were all of the 31~GHz emission proportional to H$\beta$, while
the two-component fit accounts for the presence of the
100~$\mu$m-correlated component. This is an upper limit because there
are indications (see Fig.~\ref{fig:imhelixunf} and the SED analysis in
Sec.~\ref{sec:sedmod}) that the 100~$\mu$m image is not an ideal template
for the anomalous emission at 31~GHz, so that part of it could be
mistakenly attributed to H$\beta$.

Applying $a^{II}_{\mathrm{H}\beta}$ to the observed H$\beta$ flux from
\citet{ode98} gives a 31~GHz free-free flux density of
$a^{II}_{\mathrm{H}\beta} \times F(\mathrm{H}\beta) =
2.12~10^{+12}\times3.37~10^{-13}=0.714~$Jy, or 65$\pm$19\% of
1.096~Jy, our 31~GHz flux density for the Helix (assuming a 30\%
uncertainty on the H$\beta$ measurement).




\section{SIMBA observations} \label{sec:simba}

SIMBA, at SEST, is a 37-channel bolometer array, operating at 1.2~mm
(250~GHz). The half-power beam-width of a single element is 24$''$. We
have observed the Helix on the 15 and 16 Aug. 2002, with a scanning
speed of 80~$''$~s$^{-1}$, and obtained 8 scans, $1200'' \times
1000''$ each. The SIMBA scans are reduced in the standard manner using
the MOPSI package written by Robert Zylka (IRAM, Grenoble). Flux
calibration is carried out by comparison with Uranus maps.

The 250~GHz 3-$\sigma$ upper limit after sky noise filtering,
destriping, and taking into account flux losses is 0.52~Jy, as we now
explain. The 3-$\sigma$ limit without sky noise filtering (but with
destriping) is 1.61~Jy.


\subsection{Flux loss estimates}

SEST does not operate with a chopping secondary; as a substitute sky
cancellation is obtained with channel-dependent bolometer filter
functions, which cut out low spatial frequencies in SIMBA scans. The
time sequence is multiplied by a complex bolometer filter function in
frequency space \citep{rei01}, leading to flux losses.

The $1/2$ level, for frequencies below which the 37 channel filter
functions have moduli lower than 1/2, corresponds to an average and
dispersion for all 37 filters of $6.6\pm0.4~10^{-2}$~Hz. With a
scanning speed of $80~''/$s, this is roughly equivalent to a low
spatial frequency `hole' in the $uv$-plane of radius
$170\pm10$~rad$^{-1}$. In order to estimate flux losses, we cut-out the
simulated H$\beta$ visibilities for the Helix with $uv$-radii under
170~rad$^{-1}$, and reconstruct with the maximum entropy algorithm in
AIPS++ (the fitting of simple model components is no longer a good
approximation when cutting $uv$-range). Aperture photometry of the
whole nebula and comparison with the input template shows we recover
60\% of the flux density.

\subsection{Destriping and upper limit estimates on the 250~GHz
  flux densities}

An important step in the reduction is a second sky cancellation, whose
purpose is to minimize sky noise, performed by removing correlated
signal among the 37 channels. This has been done with a search radius
of 100$''$, which in effect leaves only the outer ring of bolometer
detectors for cross-correlation. In the case of the Helix part of the
signal may have been removed - so we have also tried reducing without
sky noise removal.

The SIMBA data are affected by scanning artifacts, which appear as
noise stripes parallel to the mapping direction. We used a destriping
algorithm in order to improve the noise of the SIMBA images.  We
worked on the Fourier amplitude image of each scan, where stripes
stand out as straight lines crossing the origin (stripes have zero
frequency in one direction, and are effectively 1D point sources in
the orthogonal direction). Even faint stripes are very conspicuous in
the Fourier amplitude image, and they are easily removed by median
filtering as a function of radius, i.e. by stepping in distance from
the frequency origin and median filtering all pixels in a given
annular radial bin in the 2~D spatial frequency domain.  We flag all
visibilities whose moduli differ by more than $3\sigma$ at a given
$uv$-radius, and replace their value by an average of unflagged
visibilities in other scans.  Our destriping algorithm follows a
standard technique, detailed examples of which can be found in
\citet{eme88}, \citet{dav96}, \citet{sch98}, and its net effect is to
reduce the noise by a factor of 2.2 in the case of sky-noise-filtered
images, and a factor of 12.6 without sky-noise filtering. An example
application of our destriping algorithm can be found in \citet{hal03}.

The resulting 1$\sigma$ noise in the reduced SIMBA images is
25~mJy/beam with sky noise reduction, and 78~mJy/beam without, for a
circular beam 24$''$ FWHM.  In the Helix the bulk of optical emission
is distributed in a ring, taking about $1/3$ the solid angle of the
whole nebula, and the peaks of emission are concentrated in clumps
along the ring. We assume the bulk of emission, had it been detected,
would have fallen on 1/3 of the solid angle subtended by a uniform
disk, 660$''$ in diameter. This allows us to put 3$\sigma$ upper
limits on the integrated nebular flux density at 250~GHz of 0.52~Jy
with sky noise reduction, and 1.61~Jy without, after correcting for
60\% flux losses. The integrated flux density uncertainties were
calculated by multiplying the noise per pixel in the final SIMBA image
by $\sqrt{N\,N_\mathrm{beam}}$, in a similar way as for the CBI
photometry.


\section{Comparison with other integrated flux measurements} \label{sec:disc}

\subsection{Observed spectral energy distribution of the Helix} \label{sec:sedobs}

We have searched the literature for all existing measurements that
allow setting up a SED for the Helix. A problem with the low-frequency
data is background source contamination, which stands out in the
1.4~GHz image of \citet{rod02}. The 5~GHz data of \citet{mil74} are
distorted by a background source, as may be appreciated from Fig.~1 of
\citet{lee87}, so we take their measurement of 1.292~Jy as an upper
limit. Thus we also take the 2.7~GHz measurement of 1.27~Jy
\citep{tho70} as an upper limit. On the other hand the measurement of
0.68$\pm$0.2~Jy at 0.408~GHz, from \citet{cal82} and Calabretta (2003,
private communication), and 0.51$\pm$0.25~Jy at 0.843~GHz, from
\citet{cal85}, result after removal of background sources, and should
be accurate estimates of the nebular emission. The {\em WMAP}
first-year data release \citep{ben03} places upper limits on the Helix
flux density, which we estimate at 3$\sigma$, where $\sigma$ is the
rms point source sensitivity at the position of the Helix in 5
frequency bands.  Fig.~\ref{fig:sedhelix} summarizes the existing
measurements, together with the results from this work.

The 100$\mu$m and 60$\mu$m {\em IRAS} bands, in contrast to the
25$\mu$m and 12$\mu$m bands, are reasonably free of line emission
\citep{lee87,spe02}, and should therefore trace dust. The ISOPHOT
observations at 180~$\mu$m, 160~$\mu$m and 90$\mu$m in \citet[their
Fig.~5]{spe02} do not sample the whole nebula, but allow one to
estimate the total flux densities at 180~$\mu$m, 160~$\mu$m and
90$\mu$m by approximately scaling to the 100$\mu$m {\em IRAS}
integrated flux density\footnote{the 90$\mu$m and 180$\mu$m images of
the whole Helix available from the {\em ISO} archive seem offset from
zero, hinting at calibration problems, and would also require
scaling}.


\subsection{Problems with free-free emission} \label{sec:sedmod}

\subsubsection{Extinction in the Helix}

For free-free emission, radio continuum measurements can be  used to
predict optical recombination line fluxes, provided extinction is
known. Extinction to the Helix is very low: the highest value of the
H$\beta$ extinction coefficient reported by \citet{hen99} is $c=0.13$,
for a slit position falling on the nebular ring. Such a small value
for $c$ is within the scatter for the Balmer decrement measured by
\citet{hen99} in three slit positions.  We assume extinction is
negligible in the visible, as does \citet{ode98}.

\subsubsection{Average electron temperature}

Eq.~A8 in \citet{cap86}, also used in the PN NVSS survey of
\citet{con98}, allows one to predict an H$\beta$ flux, F(H$\beta$) in
$10^{-15}$~W~m$^{-2}$, given a thermal radio continuum flux density
$F_\nu$ in mJy at frequency $\nu$ in GHz and an electron temperature
$T_e$ in $10^{4}$~K:
\begin{equation}
F(\mathrm{H}\beta) = 0.28 \,T_e^{-0.52} \, \nu^{0.1}  F_\nu. \label{eq:hbetapredict}
\end{equation}
Our flux-loss corrected 31~GHz flux density of 1096~mJy predicts
$F_c($H$\beta) = 4.56~10^{-13}$~W~m$^{-2}$, if the electron
temperature is $T_e=9000$~K \citep{ode98,hen99}, close but in excess of
the H$\beta$ flux reported by \citet{ode98}, of
3.37~10$^{-13}$~W~m$^{-2}$. A value of $T_e=1.6\pm0.3~10^{4}~$K is
required to match the 31~GHz flux density to the H$\beta$ flux. If
$T_e=9000$~K, the expected level of free-free emission is 809~mJy,
which is significantly less than the observed 1096$\pm$100~mJy.


But, as explained in Sec.~\ref{sec:crosscor}, only part of the 31~GHz
emission is proportional to H$\beta$ (otherwise the 31~GHz and
H$\beta$ images would look the same). To match the upper limit
free-free-H$\beta$ conversion factor obtained in
Sec.~\ref{sec:crosscor}, $a^{II}_{\mathrm{H}\beta} =
(2.12\pm 0.29)~10^{12}$\,Jy~/W~m$^{-2}$, the temperature must be brought
down to $7100\pm1900$~K. The required electron temperature for the
0.408~GHz data to match the H$\beta$ flux, if dominated by optically
thin free-free emission, is $2800\pm1600~$K (assuming 30\%
uncertainties).

Surprisingly low values for the temperature of the photoionized gas in
the Helix are obtained from two independent measurements (the
31~GHz--H$\beta$ conversion factor and the 0.408~GHz flux density),
and their weighted average is $T_e = 4600\pm1200~$K.  This value is at
odds with the electron temperature derived from collisionally excited
lines (CELs), of $\sim$9000~K \citep{ode98,hen99}. However, this
discrepancy on $T_e$ has previously been reported in H\,{\sc ii}
regions and is an indication of temperature variations within the
nebulae \citep{pei67}. Our results are also reminiscent of the Balmer
jump temperature and CEL temperature discrepancy in other PNe, in
particular in M~1-42 \citep{liu01}, where the BJ temperature is
3650~K, or $\sim$5660~K less than the CEL temperature.  Other
CEL-discrepant $T_e$ diagnostics are the optical recombination line
ratios, which also give very low values, e.g. $<$2500~K in Abell 30
\citep{wes03}.

As a consistency check on Eq.~\ref{eq:hbetapredict} we compare the
value of the H$\beta$ recombination coefficient, fixed by matching the
observed H$\beta$ flux and the radio emission-measure, against the
tables in \citet{sto95}: we obtain 9.47~10$^{-14}$~cm$^{3}$s$^{-1}$,
which matches case B of \citet{bak38} values of 1.75~10$^{-13}$ at
($T_e=1000~$K, $N_e=100$~cm$^{-3}$) and 7.99~10$^{-14}$ at
($T_e=3000~$K, $N_e=100$~cm$^{-3}$), or an interpolated temperature of
2688~K, and colder yet if case A of \citet{bak38} is assumed. We
consider this value of the electron temperature is satisfactorily
close to the 3000~K derived from Eq.~A8 in \citet{cap86}.

\subsubsection{Electron temperature variations} \label{sec:tevar}

\begin{figure}
\epsscale{.5}
\plotone{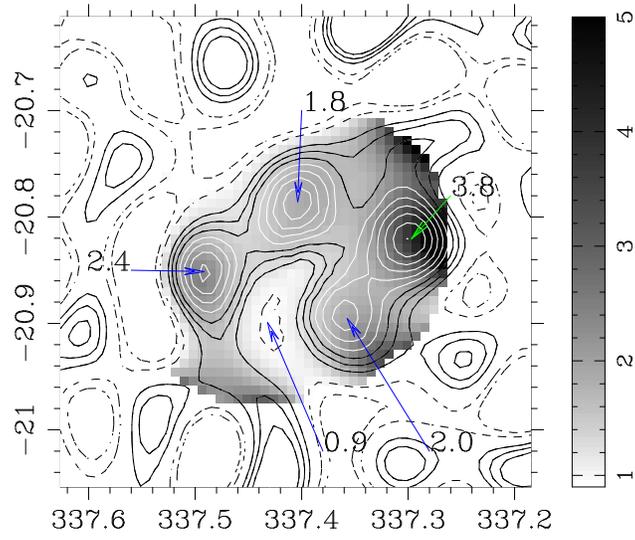}
\caption{Electron temperature map in gray scale, with an overlay of
  the anomalous 31~GHz emission from Fig.~\ref{fig:imhelixunf} in
  contours. Units are 10$^{4}$~K, and the value of $T_e$ are indicated
  at four selected positions.  \label{fig:te}}
\end{figure}

If there are temperature variations across the Helix then the
free-free emission at 31~GHz is enhanced relative to H$\beta$ in the
hot gas. We use Eq.~\ref{eq:hbetapredict} to obtain the electron
temperature map presented in Fig.~\ref{fig:te}, under the assumption
that all of the 31~GHz emission is free-free. 

The temperature structure cannot be explained in terms of
photoionization, as models predict a drop in $T_e$ with distance from
the central star \citep[their Fig.~8]{hen99}, while the temperature
peaks on the nebular ring. Trying to explain the hot spots with shock
excitation meets the difficulty that no fast wind has been detected in
the Helix \citep{cer85,cox98}. 

Although their origin is undetermined, we cannot altogether discard
that temperature fluctuations from 10\,000~K to 40\,000~K may account
for the morphological differences between H$\beta$ and 31~GHz. But
this possibility  seems rather contrived considering 1- the predicted
temperatures are higher by 10\,000~K than the [O\,{\sc iii}]
diagnostic of $9500\pm500$~K \citep{hen99}, and 2- only the Eastern
lobe clearly stands out as a hot spot, while the other regions of
anomalous emission maxima do not coincide with local $T_e$ maxima.

\subsubsection{SED modeling}

To emphasize the existence of anomalous radio emission in the Helix,
we now try modeling the SED of the Helix only taking into account
free-free and thermal vibrational emission from classical grains. We
use simple radiative transfer in a cylindrical nebula, and the {\tt
gffsub} routine in CLOUDY \citep{fer96} for the free-free gaunt
factor.  Since we need to extrapolate free-free emissivities over a
large range in frequency, we take care to validate the emissivity laws
by comparison with \citet{bec00}.  We assume the abundances of singly
and doubly ionized helium are as in \citet{hen99}, and fix the nebular
radius to 330~arcsec. The free parameters in our models are: The
proton density $N_p$, the dust temperature $T_d$, dust flux density $F_\nu$
at the reference frequency of 3000~GHz (or 100~$\mu$m), and the dust
emissivity index $\alpha$, in the form $F_\nu \propto \nu^{\alpha}
B_\nu(T_d)$. The optimization is carried-out with `Pikaia', the
genetic algorithm programmed by \citet{cha95}.

\begin{figure}
\epsscale{1.}
\plotone{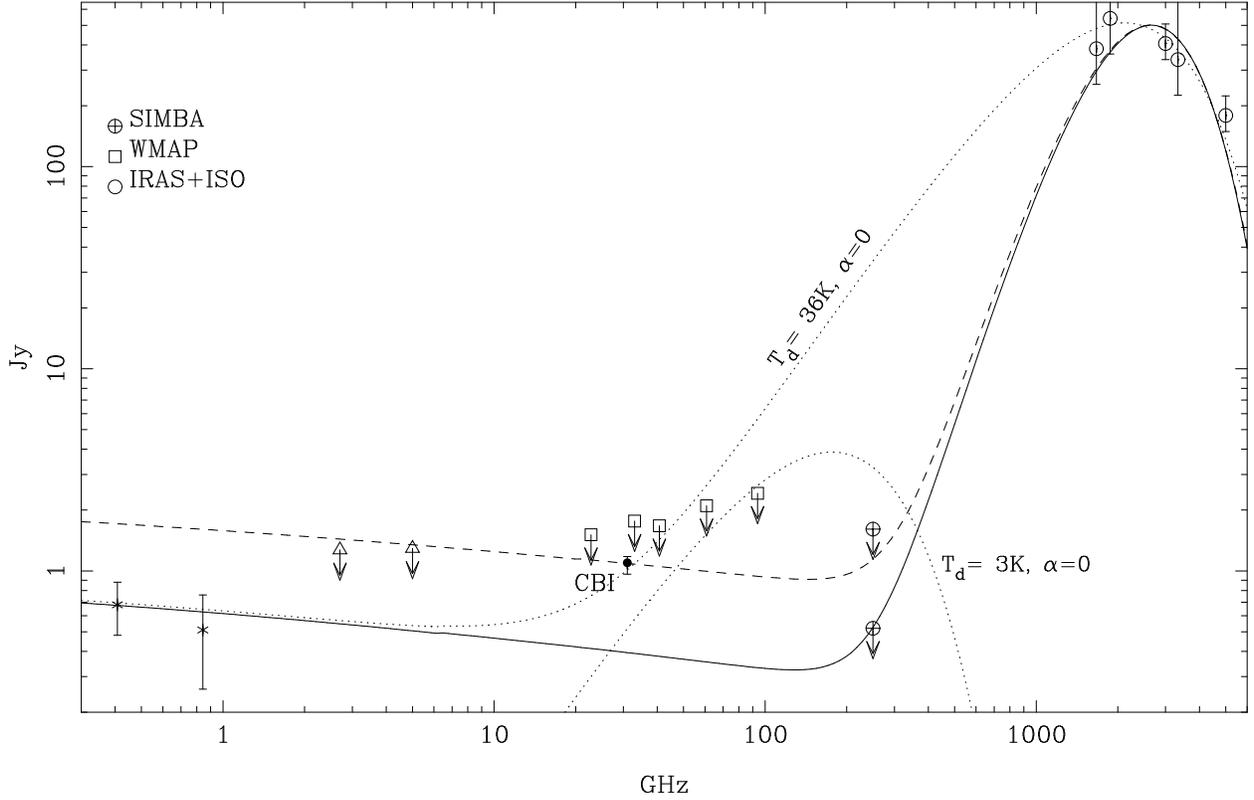}
\caption{Spectral energy distribution of the Helix. The solid line is
  a theoretical SED composed of thermal vibrational dust and free-free
  emission, where the latter uses the CLOUDY Gaunt factor and is
  designed to fit the measurement at 408~MHz \citep{cal82} and the
  {\em IRAS} and ISOPHOT data (with 20\% and 50\% errors,
  respectively).  The dashed line is a model designed to fit the CBI
  total flux density, drawn to emphasize the disagreement with low
  frequency observations.  All arrows denote upper limits, and the two
  SIMBA points correspond to the limits with (lower) and without
  (higher) sky-noise reduction.  The dotted lines show two black-body
  curves used to rule-out large grain emission as responsible for the
  31~GHz excess.  See text for information on the other
  measurements. \label{fig:sedhelix}}
\end{figure}

The resulting SEDs are shown in Fig.~\ref{fig:sedhelix}: The dashed
line fits the total CBI flux density with $T_e = 10000~$K, $T_d=24~$K,
$\alpha=2.43$, while the solid line fits the measurement at 0.408~GHz
\citep{cal82} with $T_e=3000~$K, $T_d=22~$K, $\alpha=2.8$.  Steep
values of the dust emissivity index, of order 2.5, have indeed been
reported for environments with $T_d\sim20$~K \citep{dup03}.  The best
fit dust emissivity index and temperatures are constrained by the {\em
WMAP} and SIMBA upper limit: $\alpha=0$ gives a best fit black body
emission of $\sim$1~Jy at 31~GHz for grains at 36~K, and could
otherwise account for the rise from 0.408~GHz to 31~GHz (see
Fig.~\ref{fig:sedhelix}), and the 100~$\mu$m correlation.  An
intermediate model, that would fit the far-infrared (far-IR) data and
the CBI measurement with $T_d=24~$K and $\alpha=1$, has a 250~GHz flux
density of 60~Jy, and is ruled out by the SIMBA observations.

It is apparent from Fig.~\ref{fig:sedhelix} that the 408~MHz and
31~GHz measurements cannot be reconciled under the free-free
hypothesis. The level of free-free emission is a fraction of
0.36$\pm$0.20 at 31~GHz from this spectroscopic argument.


\subsection{Ionized mass and filling factor}

An emission measure of $EM =442.6~$cm$^{-6}$~pc is required to fit the
0.408~GHz flux density, and that part of the 31~GHz flux density that
can be accounted for by free-free emission. Together with a uniform
and cylindrical model nebula 330$''$ in radius at $T_e=3000$~K, this
is a proton density of $N_p = 26$~cm$^{-3}$, and an ionized mass of
0.096~M$_{\odot}$ (with a 30\% accuracy that stems from the level of
free-free emission at 31~GHz). If all of the 31~GHz flux density were
due to bremsstrahlung and $T_e=10000$~K, then $EM=1721$~cm$^{-6}$~pc
and $N_e=52$~cm$^{-3}$.



The Helix is so diluted that it is optically thin down to 100~MHz. For
$T_e=10000$~K as well as for $T_e=3000$~K, the turnover frequency (at
which the free-free optical depth is unity), is 30~MHz. Note that if
the bulk of emission falls in 1/3 the solid angle subtended by a
uniform disk, $EM$ rises by a factor of 3, and the turnover frequency
is then about 50~MHz.

However, the Helix is renowned for its cometary knot complex. The
knots could have optically thick spectra: with a turn-over frequency
of $ \nu_T \sim 1~$GHz, the knots would be optically thin and stand
out at 31~GHz, and yet be optically thick and faint at 0.408~GHz. The
difference between the 31~GHz flux of 1096$\pm$100~mJy and
$400\pm120~$mJy, the free-free continuum at 31~GHz extrapolated from
the low-frequency data, or about $f_\mathrm{knots}=63\pm11$\% of the
31~GHz emission, would be due to the knots in their optically thin
regime. The 31~GHz--100~$\mu$m correlation would then follow by
arguing most of the dust is in the knots. But in this model the
difference between H$\beta$ and 31~GHz (both optically thin) would
stem from differential extinction of H$\beta$ by the dust in the
knots. Since the knots would not be seen in H$\beta$ to account for
the morphological differences with 31~GHz, the observed H$\beta$ flux
from the knots would vanish, $\left. F(H\beta^\mathrm{knots})
\right|_\mathrm{observed} = 0$, and the implied value of logarithmic
H$\beta$ extinction for the knot model would have to be
\begin{equation}
c=\log\left(
\frac{\left. (H\beta^\mathrm{diffuse}+H\beta^\mathrm{knots})\right|_\mathrm{dereddened}}{\left.
  H\beta^\mathrm{diffuse} \right|_\mathrm{observed}}\right)
= -\log(1-f_\mathrm{knots})= 0.43^{+0.15}_{-0.11},
\end{equation}
or $A_\mathrm{V}=0.94$ with the extinction curve from
\citet{car89}. Such an extinction coefficient would have been detected
by \citep{hen99} since their slit positions sample the nebular ring,
which is rich in knots, but they report $c<0.13$, a value which
borders their measurement uncertainty on the Balmer decrement.




Even putting aside the extinction problem, a filling factor of
$\epsilon \sim$0.5 from \citet{hen99} is far too high for the knot
model. We investigated fitting the helix SED with two free-free
components: An optically thin component and the optically thick knots,
implemented as a core-halo model. The spectral behaviour of uniform
clumps immersed in an optically thin and uniform slab is equivalent to
that of an optically thick slab inset in a larger optically thin slab,
as shown by \citet{cal91}. A turn-over frequency (at which the
free-free opacity is unity) of 7~GHz is required to accommodate both
the CBI and the 843~MHz spectral point. At $T = 10000$~K, $EM =
10^{8}~$cm$^{-6}$pc is required for $\nu_T \sim 5~$GHz. When compared
against the observed 31~GHz $EM$, this implies a filling factor
$\epsilon < 10^{-3}$. Not only is this value three orders of magnitude
lower than measured, but it is also very unusual for the ionized gas
in PNe (although common for the molecular component).

The density reported by \citet{hen99} varies between 30~cm$^{-3}$ and
upper limits of $100~$cm$^{-3}$, so the filling factor implied by the
radio data is also $\sim 0.5-1$. For the core-halo model to account
for the observed SED, the density in the clumps would have to be of
order $>10^{3}$~cm$^{-3}$, assuming the turn-over frequency of the
core is above 4~GHz and for the halo below 0.408~GHz, and using the
ratio of optically thin flux densities of both components (Mark
Calabretta, private communication). That the density diagnostics used
by \citet{hen99} miss the high density gas because of extinction in
the clumps meets the difficulty that no significant extinction is
measured with the Balmer decrement.  By comparison to other PNe, the
optical spectra of the Helix resolve very fine details relative to the
bulk nebular properties obtained from the integrated
measurements. Thus the filling factor of the Helix would be one of the
best determined, if it wasn't for the uncertainties inherent to
low-density measurements, which can only take it closer to 1.

Given the central role of the filling factor for the interpretation of
the helix SED, we searched the literature for PNe with very low value
of $\epsilon$.  The only study that has reached values of order
$\epsilon \sim 10^{-3}$ for the Helix is \citet{bof94}, with
$\epsilon=5~10^{-3}$. But we doubt the validity of this value, as we
now explain. The statistical analysis of \citet{bof94} is based on
pre-1978 data compiled by \citet{sta89}, in which the
density-diagnostic line ratios used for the Helix are at the
low-density limit of their validity. In \citet{sta89} the derived
electron density of the Helix is an upper limit only, of $N_e <
200~$cm$^{-3}$ (with substantial scatter). Moreover, \citet{bof94} do
not quote the right value from \citet{sta89} (2.8 against 2.3 for
$\log N_e$), and their H$\beta$ flux is too high by 50\% when compared
with \citet{ode98}.


\subsection{Candidate emission mechanisms at 31~GHz}

\subsubsection{The unsuitability of synchrotron or very cold grains}

What is responsible for the 31~GHz excess? Synchrotron emission is not
a suitable candidate because free-free opacity cannot account for the
turnover at $\sim$1--20~GHz, while synchrotron opacity is not expected
to be significant in a faint object, 10~arcmin in size. Moreover
synchrotron emission has never been convincingly reported in a
PN\footnote{the data mentioned by \citet{dga98} remains unpublished},
while evidence for magnetic fields is only available in very young
proto-PNe \citep{mir01}.  The record size of the Helix precludes the
 survival of a significant magnetic field, which in a
magnetized wind decreases as the inverse distance of the central star.


The 250~GHz and {\em WMAP} upper limits constrain the dust emissivity
index to a value steep enough to rule out any vibrational dust
emission below 100~GHz, for a standard grain population. But could
there be ultra-cold grains in the Helix, accounting for  half
the 31~GHz emission, or $\sim$0.5~Jy, and simultaneously reach levels
of $\sim$2~Jy at 250~GHz, thereby reconciling the SEST and CBI
measurements? The emissivity index for such grains would have to be
close to zero, to give them a broad spectrum, and their temperature
less than 3~K.  Fig.~\ref{fig:sedhelix} shows an example black-body at
3~K. Temperatures lower than 3~K have indeed been observed in the
Boomerang nebula (a post-AGB object undergoing heavy mass-loss)
through CO(1--0) absorption against the CMB \citep{sah97}.  The hot
(i.e. 20~K) and cold (i.e. 3~K) dust components have different
emissivities: The requirement $\alpha = 0$ at $\nu = 300~$GHz implies
grain sizes of order the wavelength of $\sim$1~mm, while we fit the
far-IR data with $\alpha = 2$.  For grains of a single size $a$, with
mass density of $\rho=1~$g~cm$^{-3}$,
\begin{equation} 
M_\mathrm{dust} = \frac{4}{3} \rho
  \frac{a}{Q_a(\nu)} D^2 \frac{F_\nu}{B_\nu}, 
 \end{equation} 
assuming optically thin grains, and where $F_\nu$ is the observed flux
density and $B_\nu$ is the Planck function. With the absorption cross
sections from \citet{lao93}\footnote{extracted from {\tt
www.astro.princeton.edu/$\sim$draine/}}, we have
$M^\mathrm{dust}_\mathrm{cold} = 1.1~$M$_\odot$ for 10~$\mu$m-sized
silicate grains (the largest size for which data are available), and
$M^\mathrm{dust}_\mathrm{hot} = 5.5~10^{-3}~$M$_{\odot}$ for
0.1~$\mu$m-sized grains. The implied cold dust opacity at 300~GHz is
$\chi_\nu = 1.04~$cm$^2$g$^{-1}$, in agreement with the value adopted
by \citet{jur01} for mixed-size grains with $\alpha = 0$, and with the
observed value of $\chi_\nu = 0.35~$cm$^2$g$^{-1}$ in Barnard 68
\citep{bia03}. Thus the hypothetical cold component would account for
most of the total molecular mass. Although no CO(1--0) data are
available in the literature to test for absorption, we can also
discard this cold component on the grounds that its obvious location
would be the cometary knots, but their CO temperature is
$\sim$18--40~K \citep{hug02}.

\citet{lag03} suggests that VSGs (such as PAHs or ultra-small
silicates) could reach very low temperatures after radiative
de-excitation. At 3~K the very cold VSGs would have a SED with a
significant Rayleigh-Jeans tail at 30~GHz. But we performed the same
analysis as for the very large grains and obtained that the cold dust
mass would have to be even higher than for the large grains: Using the
absorption cross sections $Q_a$ from \citet{lao93} for the smallest PAHs and
the largest silicate grains,
\begin{equation}
\left. a/Q_a(\nu) \right|_{a=3.5~10^{-4}~\mu\mathrm{m}} = 22.6
\left. a/Q_a(\nu) \right|_{a=10^{7}~\mu\mathrm{m}}.
\end{equation}

\subsubsection{New dust emission mechanisms at 31~GHz}

This leaves only two other candidates for the observed 31~GHz excess:
electric dipole emission from spinning very small dust grains
\citep[spinning dust]{dl98b}, or magnetic dipole emission from
classical grains \citep{dl99}.

VSGs are essential for the spinning dust mechanism: equipartition of
rotational energy ensures only the smallest grains, $\sim$100 atoms in
size, will rotate at 31~GHz.  But the evolutionary trend in PNe is
toward a decreasing fraction of mid-IR luminosity with size, in the
sense that mid-IR-bright PNe are the most compact, and hence the
youngest \citep{pot84}.  If VSGs always emit at mid-IR wavelengths,
then they are not expected to be present in PNe as evolved as the
Helix. \citet{lee87} have shown that the 12~$\mu$m-band IRAS flux  from
the Helix can be entirely accounted for by emission lines, suggesting
little mid-IR continuum.

The report on ISOCAM observations of the Helix \citep{cox98}
highlights the absence of VSGs in the Helix, through the absence of
characteristic PAH features and of detectable continuum at mid-IR
wavelengths.  However, this claim is unfounded because the ISOCAM CVF
pointings sample the outskirts of the nebula\footnote{namely the
western rim visible in H$_2$ emission-line images}, and cannot be
extrapolated to the whole of the Helix. Furthermore, the ISOCAM CVF
spectrum lacks the sensitivity to place useful limits on the level of
the mid-IR continuum: a noise of $\sim 1~$MJy~sr$^{-1}$ is visible in
Fig.~1 of \citet{cox98}, which when extrapolated to the whole of the
Helix is a 3~$\sigma$ limit on the total flux density of 20~Jy, while
the IRAS 12~$\mu$m flux density is 11~Jy.

\subsubsection{26--36~GHz to 100~$\mu$m conversion factors, and their spectral index}
\label{sec:corrspec}

By itself the spectrum of the Helix in the 10 CBI channels is too
noisy to place useful constraints on the spectral index. But assuming
the 100~$\mu$m HIRES map is a good template of the 31~GHz emission, we
can use the frequency dependence of the 100$\mu$m--31~GHz conversion
factors, which are determined with good accuracy, to obtain a
26--36~GHz SED for the Helix.

The determination of the 26--36~GHz spectral index derived with the
use of a template has better fractional accuracy than without. The
improved accuracy stems from assuming the morphology of the Helix is
known (fixed to that of 100~$\mu$m), so that only the
31~GHz--100~$\mu$m conversion factor is kept as a free-parameter. But
of course it must be kept in mind the 31~GHz and 100~$\mu$m maps are
not strictly identical. We provide this SED to help modeling efforts
aimed at studying the new dust emission mechanism, and as a comparison
point with \citet{fin02}. However, we caution the two-component fit
with the H$\beta$ and 100~$\mu$m templates is more significant than
the conversion factors used here, although the $a^{II}$ coefficients
are too noisy to set-up a SED.

\begin{figure}
\epsscale{0.6}
\plotone{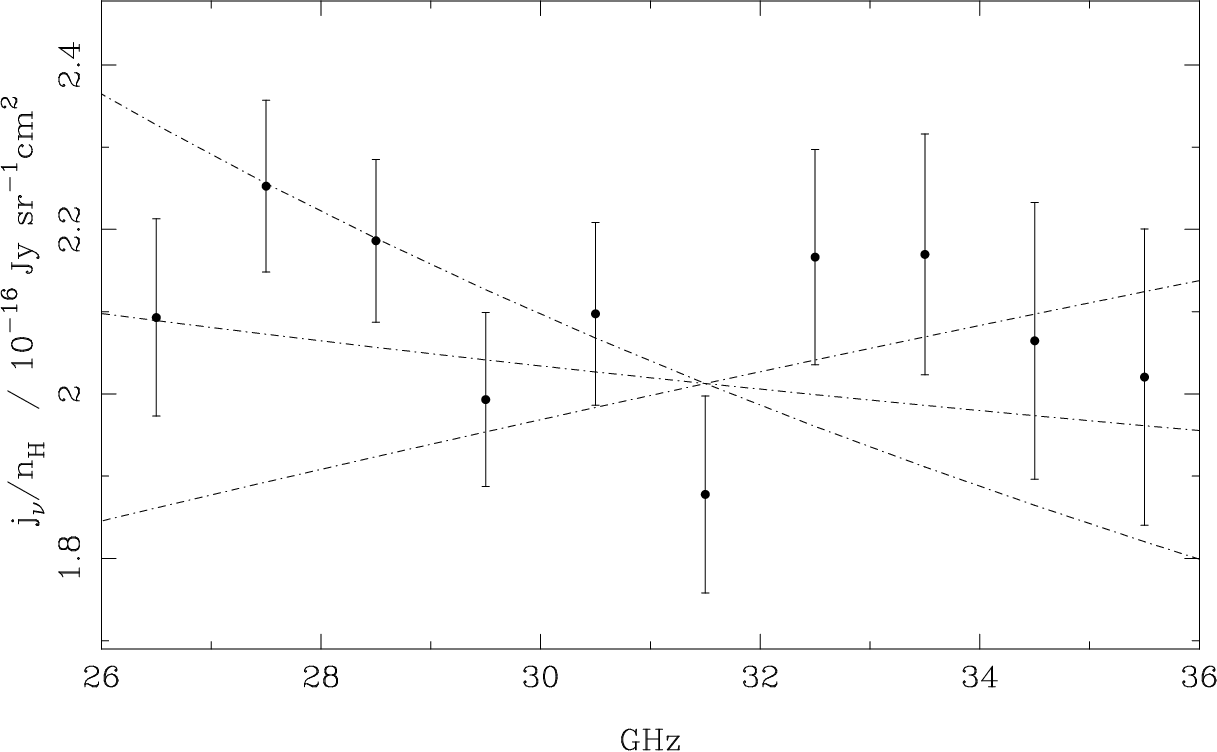}
\caption{The 26--36~GHz emissivity per nucleon, calculated by
  multiplying the peak specific intensity at 100~$\mu$m (22.9~MJy/sr
  in the CBI-simulated map) with the frequency-dependent conversion
  factors $a^{I}(100~{\mu}\mathrm{m})$. The dashed lines are power-law
  fits to the data (see text). \notetoeditor{This figure should be
  adjusted in size so as to fit in one column.}
\label{fig:ai_spec}}
\end{figure}

We obtained frequency-dependent conversion factors by
cross-correlating the real parts of the visibilities in each of the 10
CBI channels with the 100$\mu$m HIRES map (after simulating the CBI
$uv$-coverage). The spectrum for the 31~GHz emissivity per nucleon,
shown in Fig.~\ref{fig:ai_spec}, results after multiplying the
frequency-dependent $a^{I}(100~{\mu}\mathrm{m})$ with the peak
specific intensity at 100~$\mu$m (22.9~MJy/sr in the CBI-simulated
map), and dividing with an estimate of the peak nucleon column
density: $j_\nu/n_\mathrm{H}= I(31~$GHz$)/N_\mathrm{H}$, with
$N_\mathrm{H}=3.94~10^{20}~$cm$^{-2}$ for $n_\mathrm{H} =
200~$cm$^{-3}$ and a depth of 0.64~pc (approximately the diameter of
the Helix). In the Helix the molecular mass is 0.025~M$_\odot$
\citep{you99}, and the atomic mass is 0.07~M$_\odot$ \citep{rod02},
which together represent about as much as the ionized mass of
0.096~M$_{\odot}$ (this work).  Thus our estimate for the average
nucleon density along the line of sight of peak 31~GHz specific
intensity assumes about half the total mass is ionized, for which the
peak electron density is about 100~cm$^{-3}$.

The dashed lines in Fig.~\ref{fig:ai_spec} are power-law fits to the
data, with $\alpha =-0.216^{+0.67}_{-0.62}$, which is more accurate
than the photometry of Section~\ref{sec:cbiflux}, but still too noisy
to discriminate candidate emission mechanisms.  The emissivity values
in the Helix, from Fig.~\ref{fig:ai_spec}, are a factor of 2--3 higher
than the 31~GHz emissivities in LPH~201.6, from Fig.~5 in
\citet{fin02}. But we draw attention to the fact that \citet{fin02}
used an extinction-based nucleon density estimate, which is useless in
the Helix because of its anomalously high gas to dust ratio of 1000
\citep{spe02}. Considering the uncertainties involved in the
comparison, it appears $j_\nu/n_\mathrm{H}$ in the Helix and LPH~201.6
are of the same order. 

Another interesting similarity with LPH~201.6 is that the feature
which cannot be explained by free-free emission is a 5--10~GHz rise in
flux density, as it  would require unrealistic emission measures
for a diffuse H\,{\sc ii} region. This is also one of the problems
with free-free emission  in the Helix (see Sec.~\ref{sec:sedmod}).

\subsubsection{Can the 31~GHz excess be explained in terms of predicted
  magnetic dipole emissivities?}
\label{sec:Bemiss}

\citet{dl99} caution that magnetic dipole emission could significantly
contribute to the diffuse anomalous foreground if ferromagnetic grains
are abundant. Iron is strongly depleted in dust grains in most PNe
\citep{shi83,oli96}, this is perhaps also the case in the Helix. An
ionization-bounded CLOUDY model based on the line fluxes and analysis
from \citet{hen99} that reproduces [N\,{\sc
ii}]~$\lambda\lambda$6548,6584 as well as [S\,{\sc
ii}]~$\lambda\lambda$6716,6731 (with a sulfur depletion\footnote{the
difference between nebular and solar logarithmic abundance} of [S/H$]
= -1$), predicts that $F($[Fe\,{\sc ii}$]~\lambda$4300) =
$3.5~10^{-2}~F$(H$\beta$). But [Fe\,{\sc ii}]~$\lambda$4300 is absent
from the spectrum reported by \citet{hen99}, while nearby He\,{\sc
ii}~$\lambda$4026 is detected, with a flux of
$2.8~10^{-3}~F$(H$\beta$). If He\,{\sc ii}~$\lambda$4026 is at the
limit of detection, the implied iron depletion is [Fe/H]$<-1$, or less
than a tenth solar.

The emissivities shown in Fig.~\ref{fig:ai_spec} are a factor 10--20
too large to be interpreted in terms of the spinning dust or magnetic
dipole emissivities calculated by \citet{dl98b,dl99}.  However, part
of the 31GHz emission is indeed free-free, so that a 31~GHz emissivity
per nucleon derived from the 31~GHz excess is better suited to test
the model predictions.  We have estimated this excess to represent at
least 20\% of the 31~GHz emission.  This lower limit derives from the
linear combination fits to the 31~GHz visibilities using the
100~$\mu$m and H$\beta$ templates.  Applying $a^{II}(100~\mu$m) to the
peak 100~$\mu$m intensity, and using a peak nucleon column density as
above, of $N_\mathrm{H}=3.94~10^{20}~$cm$^{-2}$, we obtain
$j_\nu/n_\mathrm{H}=(4.3\pm2.3)~10^{-17}$~Jy~sr$^{-1}$~cm$^2$, now in
closer agreement with the highest magnetic dipole emissivity from
\citet{dl99}, of $8~10^{-18}~$Jy~sr$^{-1}$~cm$^2$ at 31~GHz, but still
a factor of 5 too large. 

The factor of 5 difference between the theoretical magnetic dipole
emissivities of the anomalous component could be due to different
grain temperatures: Dust grains at $T_d = 100$~K are usual in PNe,
while the ``cold neutral medium'' environment used by \citet{dl99} is
set at $T_d=18~$K, and the grain emissivities scale linearly with
$T_d$ in the Rayleigh-Jeans approximation.  The part of the grain
population that is ferromagnetic could be confined to the ionized
phase, where there is evidence for Fe depletion. This would explain
why the anomalous emission map is more compact than the 180~$\mu$m
map.  But the magnetic dipole emissivities are difficult to taylor to
the Helix because the relative number densities of ferromagnetic
grains $n_\mathrm{gr}/n_\mathrm{H}$ is undertermined: a small
population of very hot grains could also account for the observations.


%
%
%

\section{Conclusions}  \label{sec:conc}

The CBI maps and the SIMBA flux density upper limit, which represent
the highest frequency radio observations of the Helix to date, have
allowed us to investigate the nature of radio-emission mechanisms in a
diffuse circum-stellar environment. We reach the following results:

\begin{enumerate}
\item The 31~GHz map of the Helix follows a linear combination of the
100$\mu$m and H$\beta$ images.

\item The fractional level of free-free emission at 31~GHz ranges from
  $f=36\%$ to $f=80\%$ from three arguments: 1- A morphological
  argument, from the ratio of H$\beta$ coefficients in the single- and
  two-component linear combination fits, which gives an upper limit
  $f=80\pm16$\%, 2- the same morphological argument coupled with the
  observed H$\beta$ flux, which gives $f=65\pm19$\% and 3- a
  spectroscopic argument, by comparison with low-frequency
  measurements, which gives a lower limit $f=36\pm20$\%. $f=100\%$ is
  inconsistent with the 250~GHz upper limit flux density (with
  sky-noise reduction).

\item A consequence of the reduced level of free-free emission in the
  Helix is a revision of the electronic temperature for the free-free
  emitting gas to $T_e = 4600\pm1200~$K.

\item Differential extinction cannot account for the 31~GHz excess and
  morphological differences with H$\beta$ because a value of
  $A_\mathrm{V}=0.94$~mag would be required, while negligible
  extinction values of $A_\mathrm{V}<0.28$~mag have been reported
  elsewhere from optical spectra.

\item The dust-correlated 31~GHz excess over free-free emission cannot
  be explained in terms of a synchrotron component, nor with optically
  thick knots, nor in terms of ultra-cold grains of any size (as
  constrained by the 250~GHz flux density upper limit).

\item We construct a map of the anomalous 31~GHz emission by
  subtracting the H$\beta$ template image to the expected level of
  free-free emission. This map is similar to the far-IR templates, but
  is more compact, and has a much brighter Western lobe (or Western
  rim) relative to the rest of the nebula, a feature which is only
  seen in H$_2$ emission line images of the Helix.

\item Of the two models proposed by \citet{dl98b,dl99} to account for
  the anomalous foreground, we favor magnetic dipole emission from
  ferromagnetic grains over spinning dust, because of the strong iron
  depletion onto dust grains and of the suspected lack of VSGs in the
  Helix.  The inferred emissivity per nucleon for the 31~GHz excess is
  within a factor of five of the most optimistic magnetic dipole
  emissivities for the ``cold neutral medium'', which can be explained
  in terms of a dust temperature five times hotter than $T_d = 18~$K.

\item The 31~GHz--100$\mu$m conversion factor in the Helix is similar
  to the value obtained by \citet{fin02} in the diffuse H\,{\sc ii}
  region LPH~201.6. This agreement suggests the same dust emission
  mechanism is responsible for the 31~GHz excess in both objects.

\end{enumerate}




\acknowledgments

We are very grateful to the following: an anonymous referee for very
interesting comments which improved the paper and also motivated
Sec.~4.4.3 and Sec.~4.4.4; Mike Barlow for interesting suggestions,
for pointing out the role of the cometary knot complex, and for
corrections to the manuscript; Mark Calabretta for additional
information on his low-frequency measurements and interesting comments
on the free-free SED; Luis Felipe Rodriguez for providing the 1.4~GHz
image; Angela Speck for providing useful comparison images; Tom Wilson
for a critical reading. This work made use of the Southern H-Alpha Sky
Survey Atlas (SHASSA), which is supported by the National Science
Foundation. S.C. acknowledges support from FONDECYT grant
3010037. S.C.  and L.B. acknowledge support from the Chilean Center
for Astrophysics FONDAP 15010003. We gratefully acknowledge the
generous support of Maxine and Ronald Linde, Cecil and Sally
Drinkward, Barbara and Stanely Rawn, Jr., and Fred Kavli.  This work
is supported by the National Science Foundation under grant AST
00-98734.





\begin{thebibliography}{}
\bibitem[Beckert(2000)]{bec00} Beckert, T., Duschl, W.J., Mezger, P.G., 2000,
  \aap, 356, 1149
\bibitem[Baker \& Menzel(1938)]{bak38} Baker,J.G, Menzel, D.H., 1938,
  \apj, 88, 52
\bibitem[Bennett et al. (2003)]{ben03} Bennett, C. L., Halpern,
M.,Hinshaw, G., Jarosik, N., Kogut, A. , Limon, M., Meyer, S. S.,
Page, L., Spergel, D. N., Tucker, G. S., Wollack, E., Wright, E. L.,
Barnes, C., Greason, M. R., Hill, R. S., Komatsu, E., Nolta, M. R. ,
Odegard, N., Peirs, H. V. , Verde, L., Weiland, J. L., 2003, \apj,
{\em in press}, astro-ph/0302207
\bibitem[Boffi \& Stanghellini(1994)]{bof94} Boffi, F.R.,
  Stanghellini, L., 1994, \aap, 284, 248
\bibitem[Bianchi et al.(2003)]{bia03} Bianchi, S., Gon\c{c}alves, J., Albrecht,
  M., Caselli, P., Chini, R., Galli, D., Walmsley, M., 2003, \aap,
  399, L43
\bibitem[Calabretta(1982)]{cal82} Calabretta, M.R., 1982, \mnras, 199, 141 
\bibitem[Calabretta(1985)]{cal85} Calabretta, M.R., 1985,
Ph.D. thesis, "A Radio Study of Southern Planetary Nebulae",
University of Sydney, School of Physics (Dept.  of Astrophysics)
\bibitem[Calabretta(1991)]{cal91} Calabretta, M.R., 1991, Aust. J. Phys., 44, 441-58.
\bibitem[Caplan \& Deharveng (1986)]{cap86} Caplan, J., Deharveng, L., 1986,
  A\&A, 155, 297 
\bibitem[Cardelli et al.(1989)]{car89} Cardelli, J.A., Clayton, G.C., Mathis, J.S., 1989, ApJ, 345, 245
\bibitem[Cerruti-Sola \& Perinotto(1985)]{cer85} Cerutti-Sola, M.,
  Perinotto, M., 1985, \apj, 291, 237
\bibitem[Charbonneau(1995)]{cha95} Charbonneau, P., 1995, \apjs, 101, 309 
\bibitem[Condon et al.(1991]{con91} Condon, J.J., Broderick, J.J.,
Seielstad, G.A., 1991, AJ 102, 2041
\bibitem[Condon et al.(1998)]{con298} Condon, J.J., Cotton, W.D.,
  Greisen, E.W., Yin, Q.F., Perley, R.A., Taylor, G.B., Broderick,
  J.J., 1998, \aj, 115, 1693
\bibitem[Condon \& Kaplan(1998)]{con98} Condon, J.J, Kaplan, D.L.,
  1998, \apjs, 117, 361
\bibitem[Cox et al.(1999)]{cox98} Cox, P., Boulanger, F., Huggins,
P. J., Tielens, A. G. G.. M., Forveille, T., Bachiller, R., Cesarsky,
D., Jones, A. P., Young, K., Roelfsema, P. R., Cernicharo, J., 1998,
\apjl, 495, L23
\bibitem[Davies et al.(1996)]{dav96} Davies, R.D.,Watson, R.A.,
  Guti\'errez, C.M, 1996, MNRAS, 278, 925
\bibitem[Dgani \& Soker(1998)]{dga98} Dgani, R., Soker, N., 1998,
  \apjl, 499, L83
\bibitem[de\,Oliveira-Costa et al.(2002)]{deol02} de Oliveira-Costa, A,
Tegmark, M., Finkbeiner, D.P., Davies, R.D., Gutierrez, C.M., Haffner,
L.M., Jones, A.W., Lasenby, A.N., Rebolo, R., Reynolds, R.J., Tufte,
S.L., Watson, R.A., 2002, \apjl, 567, 363
\bibitem[de\,Oliveira-Costa et al.(1999)]{deol99} de Oliveira-Costa,
A., Tegmark, M., Gutiérrez, C.M., Jones, A.W., Davies, R.D., Lasenby,
A.N., Rebolo, R., Watson, R.A., 1999, \apjl, 527, 9
\bibitem[Draine \& Lazarian(1998)]{dl98a} Draine, B.T., Lazarian, A., 1998,
\apjl, 494, L19
\bibitem[Draine \& Lazarian(1998)]{dl98b} Draine, B.T., Lazarian, A., 1998,
\apj, 508, 157
\bibitem[Draine \& Lazarian(1999)]{dl99} Draine, B.T., Lazarian, A., 1999,
\apj, 512, 740
\bibitem[Dupac et al.(2003)]{dup03} Dupac, X., Bernard, J.-P., Boudet, N.,
  Giard, M., Lamarre, J.-M., M\'eny, C., Pajot, F., Ristorcelli, I.,
  2003, proceeding ``Multi-Wavelength Cosmology'' conference held in
  Mykonos, Greece, June 2003, ed. Kluwer, {\em in press}
\bibitem[Emerson \& Gr\"ave(1988)]{eme88}  Emerson, D.T., Gr\"ave, R., 1988, A\&A 190, 353
\bibitem[Ferland(1996)]{fer96} Ferland G.J., 1996, {\em Hazy, a brief
  introduction to Cloudy}, University of Kentucky Department of
  Physics and Astronomy Internal Report (Hazy)
\bibitem[Finkbeiner et al.(2002)]{fin02} Finkbeiner, D.P., Schlegel, D.J.,
Frank, C., Heiles, C., 2002, \apj, 566, 898
\bibitem[Gaustad et al.(2001)]{gau01} Gaustad, J.E., McCullough, P.R.,
Rosing, W., Van Buren, D., 2001, \pasp, 113, 1326
\bibitem[Hales et al.(2003)]{hal03} Hales, A., Casassus, S., Alvarez,
  H., May, J., Bronfman, L., Readhead, A.C., Pearson, T.J., Mason,
  B.S., Dodson, R., 2003, \apj, {\em in preparation}
\bibitem[Harris et al.(1997)]{har97} Harris, H.C, Dahn, C.C., Monet,
  D.G., Pier, J.R.,  1997, IAU symp. 180, Planetary Nebulae,
  ed. H.J. Habing, H.J.G.L. Lamers (Dordrecht: Kluwer), 40 
\bibitem[Henry et al.(1999)]{hen99} Henry, R.B.C., Kwitter, K.B.,
  Dufour, R.J., 1999, \apj, 517, 782 
\bibitem[Huggins et al.(2002)]{hug02} Huggins, P.J., Forveille, T.,
  Bachiller, R., Cox, P., Ageorges, N., Walsh, J.R., 2002, \apjl, 573,
  L55.
\bibitem[Jura et al.(2001)]{jur01} Jura, M., Webb, R.A., Kahane, C.,
  2001, \apjl, 550, L71
\bibitem[Kogut et al.(1996)]{kog96} Kogut, A., Banday, A.J., Bennett,
  C.L., Gorski, K.M., Hinshaw, G., Reach, W.T., 1996, \apj, 460, 1
\bibitem[Laor \& Draine(1993)]{lao93} Laor, A,  Draine, B.T., 1993, \apj, 402,441
\bibitem[Lagache(2003)]{lag03} Lagache, G.,2003, \aap, {\em in press}
\bibitem[Lang(1980)]{lan80}Lang,K.R., 1980, ``Astrophysical Formulae'', Springer-Verlag
\bibitem[Leene \& Pottasch(1987)]{lee87} Leene, A., Pottasch, S.R.,
  1987, \aap, 173, 145
\bibitem[Leitch et al.(1997)]{lei97} Leitch, E.M., Readhead, A.C.S., Pearson,
  T.J., Myers, S.T., \apjl 486, L23
\bibitem[Lemke et al.(1996)]{lem96} Lemke, D. et al 1996, \aap 315, L64
\bibitem [Liu et al.(2001)]{liu01} Liu, X.-W., Luo, S.-G., Barlow,
M. J., Danziger, I. J., Storey, P. J., 2001, \mnras, 327, 141
\bibitem[Mason et al.(2003)]{mas03} Mason, B. S. , Pearson, T. J.,
 Readhead, A. C. S. , Shepherd, M.C.  Sievers, J. L., Udomprasert,
 P. S., Cartwright, J. K., Farmer, A. J., Padin, S., Myers, S. T.,
 Bond, J. R., Contaldi, C. R., Pen, U.-L., Prunet, S., Pogosyan, D.,
 Carlstrom, J. E., Kovac, J., Leitch, E. M., Pryke, C., Halverson,
 N. W., Holzapfel, W. L., Altamirano, P., Bronfman, L., Casassus, S.,
 May, J., Joy, M., 2003, \apj, 591, 540
\bibitem[McCullough \& Chen(2002)]{cul02} McCullough P.R., Chen, R.R., 2002,
  \apjl, 566, 45
\bibitem[Milne \& Aller(1974)]{mil74} Milne D.K., Aller, L.H., 1974,
  Galactic Radio Astronomy, eds. F.J.Kerr, S.C., Simonson, p. 411 
\bibitem[Milne \& Aller(1982)]{mil82} Milne, D.K., Aller, L.H., 1982,
  \aaps, 50, 209
\bibitem[Miranda et al.(2001)]{mir01} Miranda, L.F., Gomez, Y.,
  Anglada, G., Torrelles, J.M., 2001, \nat, 414, 284
\bibitem[Nyman et al.(2001)]{nym01} Nyman, L.\AA., Lerner, E.A.,
  Nielbock, M., et al., 2001, The Messenger (ESO), 106, 40
\bibitem[O'Dell(1998)]{ode98}O'Dell, C.B., 1998, \aj, 116, 1346 
\bibitem[Oliva et al.(1996)]{oli96} Oliva, E., Salvatti, M., Moorwood,
  A.F.M., Marconi, A., 1994, \aap, 288, 457
\bibitem[e.g. Osterbrock (1989)]{ost89} Osterbrock D.E., 1989,
``Astrophysics of Gaseous Nebulae and Active Galactic Nuclei'',
W.H.Freeman and Company, San Francisco
\bibitem[Padin et al.(2002)]{pad02} Padin, S., et al, 2002, \pasp, 114, 83
\bibitem[Page et al.(2003)]{pag03} {Page}, L., {Barnes}, C.,
	{Hinshaw}, G., {Spergel}, D.~N., {Weiland}, J.~L., {Wollack},
	E., {Bennett}, C.~L., {Halpern}, M., {Jarosik}, N.,
	{Kogut}, A., {Limon}, M., {Meyer}, S.~S., {Tucker},
	G.~S., {Wright}, E.~L., 2003, \apjs, 148, 39
\bibitem[Peimbert(1967)]{pei67} Peimbert, M., 1967, \apj, 150, 825
\bibitem[Pottash et al.(1984)]{pot84} Pottasch, S.R., Baud, B.,
Beintema, D., Emerson, J., Harris, S., Habing, H.J., Houck, J.,
Jennings, R., Marsden, P., 1984, \aap, 138, 10
\bibitem[Readhead \& Pearson (2003)]{rea03} Readhead, A.C.R, Pearson, T.J.,
  2003, in Carnegie Observatories Astrophysics Series, Vol. 2: Measuring
  and Modeling the Universe, ed. W. L. Freedman (Cambridge: Cambridge
  Univ. Press);
\bibitem[Reichertz et al.(2001)]{rei01} Reichertz, L. A., Weferling,
B., Esch, W., Kreysa, E., 2001, \aap, 379, 735
\bibitem[Rodriguez et al.(2002)]{rod02} Rodriguez, L.F., Goss, W.M.,
  Williams, R., 2002, \apj, 574, 179
\bibitem[Sahai \& Nyman(1997)]{sah97} Sahai, R., Nyman, L-\AA, \apjl, 487, L155
\bibitem[Schlegel et al.(1998)]{sch98} Schlegel, D.J., Finkbeiner,
D.P., Davis, M., 1998, ApJ 500, 525
\bibitem[Shepherd(1997)]{she97} Shepherd, M.C., 1997, in Astronomical
  Data Analysis Software and Systems VI, ed. G~Hunt \& H.E.~Payne, ASP
  conference series, v125, 77-84 ``Difmap: an interactive program for
  synthesis imaging''.
\bibitem[Shields(1983)]{shi83} Shields, G.A., 1983, in Flower D.R.,
  ed., IAU Symp. 103, Planetary Nebulae, Kluwer, Dordrecht, p. 259
\bibitem[Speck et al.(2002)]{spe02} Speck, A.K., Meixner, M., Fond, D.,
McCullough, P.R., Moser, D.E., Ueta, T., 2003, \aj, 123, 346
\bibitem[Stanghellini \& Kaler(1989)]{sta89} Stanghellini, L., Kaler,
  J.B.,  1989, \apj, 343, 811
\bibitem[Storey \& Hummer(1995)]{sto95} Storey, P.J., Hummer, D.G.,
  1995, \mnras, 272, 41
\bibitem[Thomasson \& Davies (1970)]{tho70} Thomasson, P., Davies,
  J.G, 1970, \mnras, 150, 359
\bibitem[Wesson et al.(2003)]{wes03} Wesson, R., Liu, X.-W., Barlow, M. J., 2003, MNRAS, 340, 253
\bibitem[Young et al.(1999)]{you99} Young, K., Cox, P., Huggins,
  P.J., Forveille, T., Bachiller, R., 1999, \apj, 522, 387
\bibitem[Zijlstra et al.(1989)]{zij89} Zijlstra, A.A., Pottasch,
  S.R., Bignelli, C., 1989, \aaps, 79, 329



\end{thebibliography}
\end{document}